\documentclass[journal=acscii,manuscript=article]{achemso}

\usepackage[version=4]{mhchem} 
\usepackage{amsmath}
\usepackage{amssymb}
\usepackage{graphicx}
\usepackage{color}
\usepackage{tablefootnote}
\usepackage{pdfpages}

\newcommand*{\rtvec}[1]{\mathbf{#1}}

\author{Zeeshan Ahmad}
\affiliation{Department of Mechanical Engineering, Carnegie Mellon University, Pittsburgh, Pennsylvania 15213, USA}
\author{Zijian Hong}
\affiliation{Department of Mechanical Engineering, Carnegie Mellon University, Pittsburgh, Pennsylvania 15213, USA}
\author{Venkatasubramanian Viswanathan}
\affiliation{Department of Mechanical Engineering, Carnegie Mellon University, Pittsburgh, Pennsylvania 15213, USA}
\alsoaffiliation[chem]{Department of Chemical Engineering, Carnegie Mellon University, Pittsburgh, Pennsylvania 15213, USA}
\email{venkvis@cmu.edu}

\title[paper title]
  {Design rules for liquid crystalline electrolytes for enabling dendrite-free lithium metal batteries}


\begin{document}

\begin{tocentry}
\includegraphics[width=8.4cm]{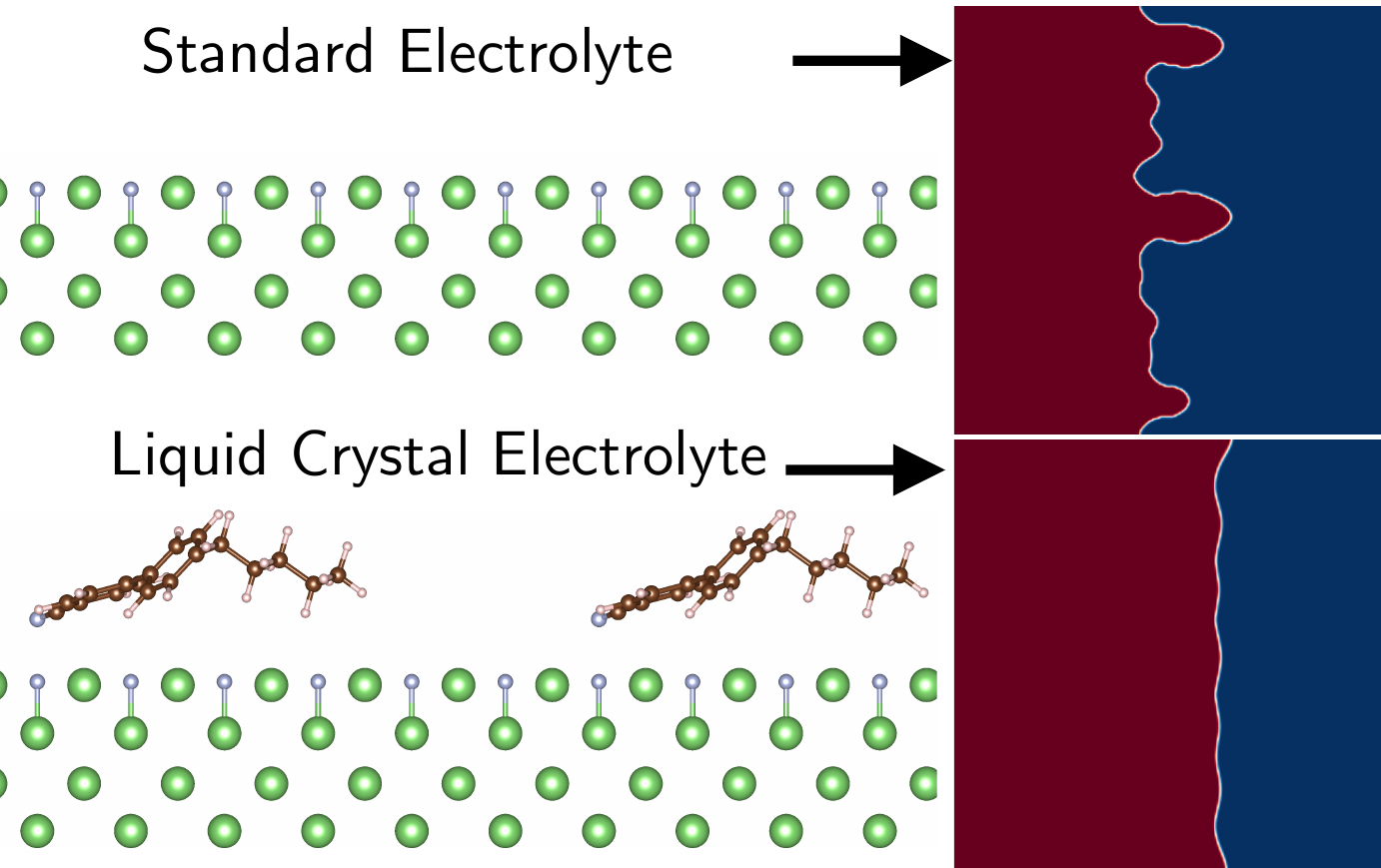}




Synopsis: We propose liquid crystals with high anchoring energies can suppress dendrite growth in Li-ion batteries with a Li metal anode.
\end{tocentry}

\begin{abstract}
Dendrite free electrodeposition of lithium metal is necessary for the adoption of high energy density rechargeable lithium metal batteries. Here, we demonstrate a new mechanism of using a liquid crystalline electrolyte to suppress dendrite growth with a lithium metal anode. A nematic liquid crystalline electrolyte modifies the kinetics of electrodeposition by introducing additional overpotential due to its bulk distortion and anchoring free energy. By extending the phase-field model, we simulate the morphological evolution of the metal anode and explore the role of bulk distortion and anchoring strengths on the electrodeposition process. We find that adsorption energy of liquid crystalline molecules on lithium surface can be a good descriptor for the anchoring energy and obtain it using first-principles density functional theory calculations.  Unlike other extrinsic mechanisms, we find that liquid crystals with high anchoring strengths can ensure smooth electrodeposition of lithium metal, thus paving the way for practical applications in rechargeable batteries based on metal anodes.
\end{abstract}

\section{Introduction}
The development of high energy density rechargeable batteries is essential to meet the goals of decarbonization through electrification of transportation~\cite{sripad2017lcv,moore2014aviation} and storage of renewable technologies. The ideal anode material for lithium (Li) ion batteries is Li metal with a specific capacity over ten times that of the currently commercialized graphite anode~\cite{XuLi2014,xin-bing2017lithium}. It has, however, been plagued by issues related to uneven electrodeposition during charging leading to growth of dendrites and loss of Coulombic efficiency. This problem is more prominent during fast charging which is a necessity for practical applications of Li-ion batteries in electric vehicles and flights~\cite{Ahmed2017enabling}. 
Numerous approaches for suppressing dendritic growth have been proposed. These include the use of artificial solid electrolyte interphase (SEI)~\cite{liu2015artificial,yan2014ultrathin,liu2017artificial}, additives in liquid electrolytes~\cite{aurbach1996comparative,hirai1994effect,ding2013electrostatic,qian2015high, suo2013solvent,lu2014stable,zhang2017FEC}, surface nanostructuring~\cite{Wang17,zhang2014nanorod}, solid polymer or inorganic electrolytes~\cite{khurana2014suppression, Li2018suppressing, balsara2012-modadh,Barai2017lithium, li2014lipon,Li2019electroconvection, fu2019universal}.  Among the many approaches, mechanical suppression of dendrite growth through stresses at solid-solid interfaces provides design principles for the desired mechanical properties for polymer~\cite{Monroe2005Impact} and inorganic solid electrolytes~\cite{ahmad2017stability, ahmad2017-anisotropy}.  These analyses suggest that polymers with high shear modulus and ceramics with low shear modulus can lead to stable electrodeposition.  Generally, most polymers tend to be soft while ceramics tend to be hard and hence finding ceramic or polymer materials that satisfy the stability criterion has proved challenging~\cite{Ahmad2018machine}.

Liquid crystalline materials offer an interesting new avenue to suppress dendritic growth through additional energetic contributions that emerge due to distortion and anchoring.  These energies originate from the tendency of the anisotropic molecules to reorient and align resulting in an ordered arrangement~\cite{degennes1974liquid}. As compared to the other dendrite suppression methods, liquid crystals are easy to synthesize, manufacture and integrate into batteries, while offering the potential to spontaneously suppress dendrites without external forces (e.g. stack pressure). Liquid crystal surface properties like anchoring strengths are of importance in opto-electronic applications like liquid crystal displays, lithography and molecular electronics~\cite{Meier1975applications}. Solid substrates in contact with liquid crystals tend to uniformly align their molecules~\cite{degennes1974liquid} which led to the advent of liquid crystal displays, promoting research on elaborate characterization of molecular orientation with different substrates and the effects of surface treatment methods like rubbing, coating and surface active agents on surface alignment~\cite{Castellano1983surface, Kahn1973surface, Zocher1928surface, Yokoyama1988surface}. Recently, Li-containing liquid crystalline materials have been developed as electrolytes possessing high ionic conductivity~\cite{Kato2010from}. Engineered liquid crystalline materials are attractive candidates as electrolytes due to the presence of organized 1D, 2D or 3D pathways for ion conduction~\cite{Shimura2009electric, Kishimoto2005nano,  Kato2010from, Gadjourova2001ionic}. Further, high transference number, low-cost bulk manufacturing, low-flammability and wider temperature window may be achieved in conjunction with fast ionic transport in the crystalline phase compared to amorphous phase~\cite{Gadjourova2001ionic, Imrie1999ion, zimmerman2017solid, zimmerman2017solid2}. ~\citealt{Kerr2009new} and ~\citealt{Kato2015liquid} obtained ionic conductivities in the range $10^{-6}-10^{-3}$ S/cm at room temperature. \citealt{Kato2015liquid} further demonstrated reversible charge-discharge cycling with \ce{LiFePO4} and \ce{Li4Ti5O2} cathodes and Li metal anode. While ionic conductivity and voltage stability of liquid crystalline electrolytes are promising, natural questions that emerge are 1) whether liquid crystalline electrolytes in contact with a metal anode can suppress dendrites and 2) what liquid crystal properties affect the suppression. In this work, we combine phase-field simulations and density functional theory (DFT) calculations to study electrodeposition with a liquid crystalline electrolyte.  We simulate the electrodeposition of a metal anode in the presence of the liquid crystalline electrolyte using a phase-field model by including the effects of bulk distortion and anchoring energies on the kinetics. We find that liquid crystalline electrolytes with sufficient anchoring strengths at the interface with the metal anode can ensure smooth electrodeposition and greatly suppress dendrite growth. We quantify dendrite suppression using three metrics based on the shape and location of the interface and its growth with time. Based on the analysis of these metrics, we provide design rules for material architectures that can suppress dendritic growth.  Having identified the design rules, using DFT calculations, we identify descriptors for the anchoring energy and stability at the cathode-liquid crystalline electrolyte interface.  We find that adsorption strength of liquid crystalline molecules at the lithium metal surface is a good descriptor to describe anchoring strength.  We also screen for the highest occupied molecular orbital (HOMO) level for the liquid crystalline molecules to get an estimate of their oxidative stability. Given the very large design space of possible liquid crystalline electrolytes, we believe that these descriptors are stepping stones towards a larger high-throughput search over the design space.

\section{Liquid Crystalline Electrolyte Design Considerations}
Liquid crystalline phases are commonly found in materials composed of anisotropic molecules which interact with one another~\cite{degennes1974liquid}. In the simplest liquid crystalline phase called the nematic phase, molecules tend to orient parallel to each other giving rise to orientational order but no long range positional order.   Liquid crystalline electrolytes typically consist of liquid crystalline phase mixed with a conventional organic solvent and lithium containing electrolyte salt. Examples include a rod-like molecule having a mesogenic unit and an alkyl chain terminating in a carbonate moiety and lithium bis(trifluoromethylsulfonyl)imide (LiTFSI) used by ~\citealt{Kato2015liquid} and amphiphilic $\beta$-cyclodextrins with LiTFSI proposed by ~\citealt{Champagne2019liquid}

For use of these liquid crystalline in a battery, an electrolyte must satisfy numerous properties simultaneously.  It must possess sufficient electrochemical stability at the anode and cathode chosen, high ionic conductivity, low electronic conductivity, thermal stability and good adhesion (wetting) at the interface~\cite{Xu2004nonaqueous}. 

\begin{figure}[htbp]
\centering
\includegraphics[scale=0.52]{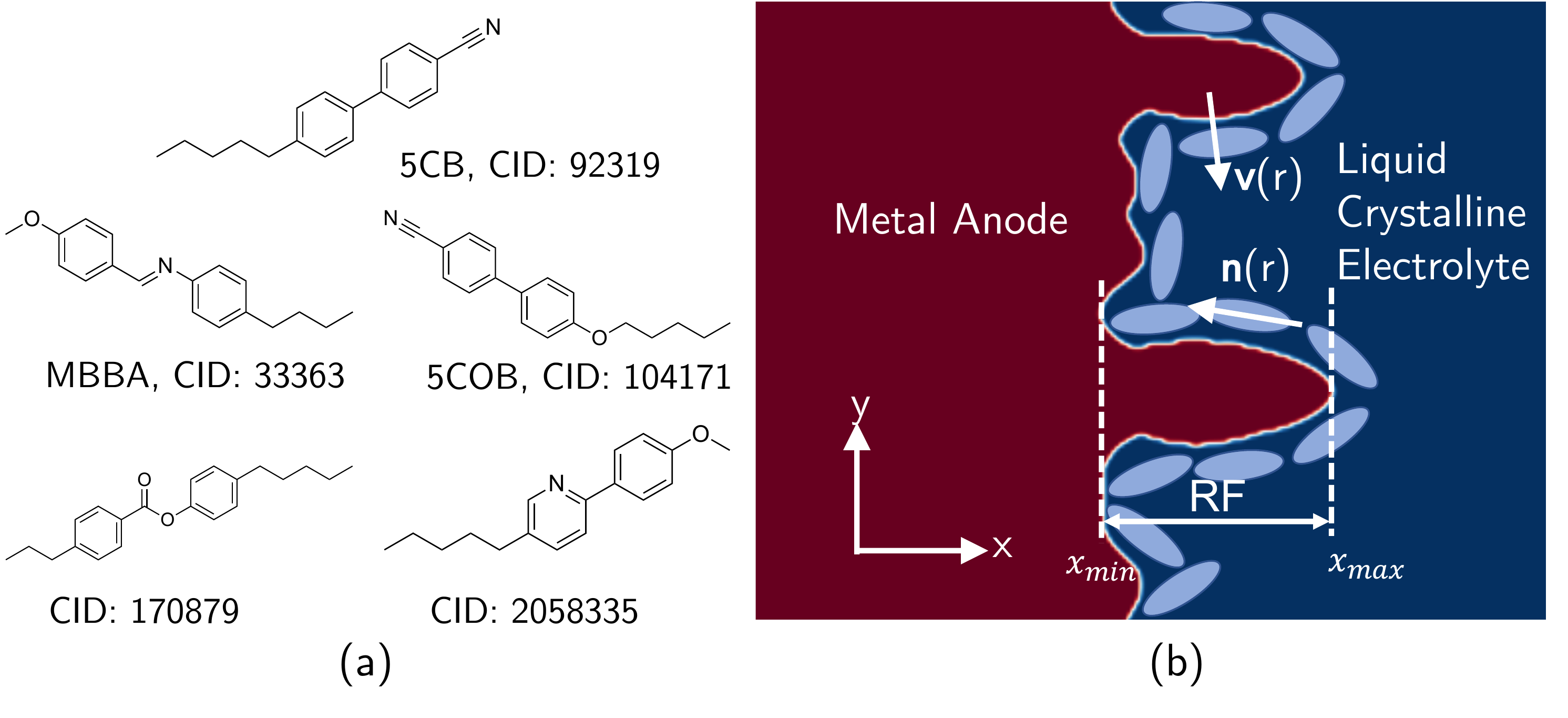}
\caption{\label{fig:schematic}
(a) Structures of some common liquid crystal molecules with their PubChem CIDs. (b) Schematic of an interface between metal electrode and liquid crystalline electrolyte. $\rtvec{n}(r)$ is the director field of the liquid crystal. The liquid crystal molecules (size exaggerated) orient along the surface of the electrode due to anchoring energy.}
\end{figure}

The possible design space of liquid crystalline electrolytes is large~\cite{Kato2010from}.  For the liquid crystalline molecule chosen, properties can be be tuned through the length of the backbone chain, functional end groups and salt molarity~\cite{Gray1971molecular, Jerome1991surface}. The ionic conductivity could be tuned by modifying the fraction of Li salt used~\cite{Kato2015liquid}.  Common liquid crystals include $n$-cyanobiphenyls ($n$CB, $n$ being the length of alkyl chain), N-(4-Methoxybenzylidene)-4-butylaniline (MBBA) \&  4-(4-pentoxyphenyl)benzonitrile (5COB)  and are shown in Fig.~\ref{fig:schematic}a.  MBBA and 5CB  are most commonly used for physical and electrooptical investigations of nematic liquid crystals.

Most liquid crystals are electronically insulating but conductive ones have been designed with 1D or 2D conductivity through doping~\cite{Zhang2015high, Boden1988one}. The conductivity of disc-shaped solid-like 2,3,6,7,10,11-hexakis-
(hexyloxy)triphenylene (HAT6) has been shown to increase from below $10^{-9}$ S m$^{-1}$ to $10^{-3}$ S m$^{-1}$ on doping with the Lewis acid \ce{AlCl3} with p-type conduction~\cite{Boden1988one}. In general, liquid crystals with low oxidation (reduction) potentials can be doped with electron acceptors (donors) to increase electronic conductivity~\cite{Boden1994first}. Rod-like molecules like the ones mentioned above are generally electronically insulating and only show electronic conduction only under very high purity~\cite{Hanna2012charge}.

Nematic liquid crystals undergo a first order nematic to isotropic transition as the temperature is increased to the transition temperature $T_{\text{NI}}$ with a sudden decrease in the order parameter. The transition temperature can be increased, for example, by increasing the length $n$ of the alkyl chain in $n$CB while also exhibiting the odd-even effect~\cite{Gray1971molecular,Tiberio2009}. The transition is reversible which ensures that the properties can be recovered in case the temperature increases during device operation. Liquid crystalline macromolecular structures or polymers exhibit better chemical stability and low flammability~\cite{Poletto2016design,zimmerman2017solid}. 

All of this points to the promise of using liquid crystalline electrolyte in battery systems.  In our design space analysis, we focus on electrochemical stability and desired properties required for suppressing dendrites at the lithium metal-liquid crystalline electrolyte interface.  We use phase-field simulations combined with density functional theory (DFT) to calculate these properties.  In the next section, we formulate electrochemical kinetics in the presence of a liquid crystalline electrolyte.

\section{Electrodeposition Kinetics}
We discuss below the reformulation of electrochemical kinetics in the presence of liquid crystalline electrolytes, in the context of a phase-field model.   Fig. \ref{fig:schematic}b shows a schematic of a metal anode in contact with a nematic liquid crystalline electrolyte. The average orientation of the molecules is given by a unit vector $\rtvec{n}$ which is called the director. In a distorted liquid crystal, the director will vary with space $\rtvec{n}=\rtvec{n}(r)$. The bulk distortion free energy of the liquid crystal can be written in terms of the director field $\rtvec{n}$ using three elastic constants corresponding to splay, twist and bend deformations~\cite{degennes1974liquid}.

In this work, we use the one-constant approximation~\cite{degennes1974liquid} with elastic constant $K$ which gives a simple form of the distortion free energy:
\begin{equation}
    F_d =\int \text{d}V \frac{1}{2}K(\nabla \rtvec{n}\cdot \nabla \rtvec{n})
\end{equation}
This equation is widely used for simulating liquid crystal behavior due to its simple form and gives valuable insights on distortions in nematics. Berreman used a similar form of the distortion free energy to explain the orientation of nematics in contact with a solid surface and grooved surface~\cite{Berreman1972solid, Berreman1973alignment}. In addition to bulk distortion free energy, existence of an interface with the Li metal will result in certain preferred directions for the director of the molecules in contact~\cite{Jerome1991surface}, called the easy axes. The easy axes can be, for example, crystallographic directions for an interface with a single crystal or at a certain angle to the surface in the case of MBBA free surface~\cite{degennes1974liquid, Bouchiat1971molecular}. This results in the so called strong anchoring of the nematic phase at an interface. Here we use the form of anchoring energy derived from the expression proposed by Rapini and Papoular~\cite{Stelzer1997rapini, Rapini1969distortion}

\begin{eqnarray}\label{eq:anch}
F_{\text{anch}}=\int \text{d}S\frac{1}{2} W(\rtvec{n}\cdot \rtvec{v})^2
\end{eqnarray}
In this equation, $\rtvec{v}$ is the normal to the interface between the electrode and the electrolyte, $W$ is the anchoring strength and the integration is performed over the interface area (See Supporting Information for derivation). This energy favors alignment of the liquid crystal molecules along the tangent to the interface as shown in Fig. \ref{fig:schematic}b.

Since the morphological evolution of the metal anode under electrodeposition involves moving interfaces, phase-field models are ideal to treat this problem without having to track the actual position of the interface~\cite{Karma1998quantitative, Chen2002phase}. A phase-field model is an efficient simulation tool to obtain mesoscale insights on phase transitions, transformations and microstructure evolution \cite{Chen2002phase, Boettinger2002phase, Warren2003extending, karma2001phase}. Previously, several phase-field simulation models have been developed for obtaining a quantitative understanding of dendrite growth in Li-ion batteries \cite{Chen2015modulation, Cogswell2015quantitative, Ely2014phase, bazant2013theory, Hong2018phase, Hong2019prospect}. Here we use the fully open source MOOSE (Multiphysics Object-Oriented Simulation Environment) framework~\cite{Gaston2009moose} to solve the phase-field equations~\cite{Schwen2017rapid}.

Our phase-field model uses the grand potential functional to generate the phase-field equations~\cite{Plapp2011unified,Hong2018phase}. This formulation permits the use of a larger interface thickness $\delta$ for computational convenience, much greater than the physical width of the interface while eliminating non-physical effects. The phase-field variable $\xi$ is a nonconserved order parameter whose value is 1 for the metal anode (solid) and 0 for the electrolyte (liquid). We use the Butler Volmer - Allen Cahn reaction (BV-ACR) equation for the evolution of the phase-field variable $\xi$ with time $t$ \cite{bazant2013theory, Chen2015modulation, Hong2018phase}:

\begin{eqnarray}\label{eq:evolxi}
\frac{\partial \xi}{\partial t} = -\mathcal{M}_{\sigma} [g'(\xi) -\kappa \nabla^2 \xi ] - \mathcal{M}_{\eta} h'(\xi) f(\eta) - q 
\end{eqnarray}
where $\mathcal{M}_{\sigma}$ and $\mathcal{M}_{\eta}$ are the interfacial mobility and electrochemical reaction kinetics coefficient respectively. $g(\xi)=b\xi^2 (1-\xi)^2$ is a double-well function where $b$ is related to the switching barrier and $\kappa$ is the gradient energy coefficient. The surface energy $\gamma$ and interface thickness $\delta$ can be used to obtain the values of $b=12\gamma/\delta$ and $\kappa=3\delta \gamma/2$ \cite{Cogswell2015quantitative, Cahn1959free}. $h(\xi ) = \xi^3(6\xi^2 -15 \xi +10)$ is the interpolation function for $\xi$. A Langevin noise term $q$ is added to the equation to account for thermal and structural fluctuations. $f(\eta)$ is related to the kinetics of electrodeposition in terms of the total overpotential $\eta$ at the interface. The overpotential and hence the kinetics is modified by the liquid crystal due to a change in equilibrium potential difference between the electrode and the electrolyte. Let $\eta_0$ be the overpotential without the liquid crystal and $\eta_{LC}$ be the additional overpotential due to the liquid crystal i.e. $\eta_{\text{LC}}=\eta - \eta_0$. The charge transfer coefficient for the liquid crystal overpotential $\eta_{\text{LC}}$ will in general be different from that for $\eta_0$~\cite{diggle1969zinc, Monroe2004Effect, ahmad2017stability, McMeeking2019metal, Jana2017lithium}. Assuming a Butler Volmer equation with charge transfer coefficients $\alpha_c$, $\alpha_a$ for the cathode and anode, and a cathodic charge transfer coefficient $\alpha_d$ for the liquid crystal overpotential, we obtain:

\begin{eqnarray}
f(\eta) = \frac{c_{\text{Li}^+}}{c_0} \exp\left[ -\frac{\alpha_d n\mathcal{F}\eta_{\text{LC}}}{RT} \right] \exp\left[ -\frac{\alpha_c n\mathcal{F}\eta_0}{RT} \right]\nonumber\\
- \exp \left[ \frac{(1-\alpha_d) n\mathcal{F}\eta_{\text{LC}}}{RT} \right] \exp \left[ \frac{\alpha_a n\mathcal{F}\eta_0}{RT} \right]
\end{eqnarray}

Parameters $n$, $\mathcal{F}$, $R$ and $T$ are the number of electrons transferred, Faraday's constant, gas constant and temperature respectively. The overpotential $\eta$ can be written in terms of the actual and equilibrium potential at the anode ($\phi_e$) and the electrolyte $\phi$ as $\eta=(\phi_e - \phi) - (\phi_e - \phi)_{\text{eqbm}}$. $c_{\text{Li}^+}$ is the mole fraction of Li and $c_0$ is its standard mole fraction. The charge transfer coefficient $\alpha_d$ can be associated with mechanical deformation and heterogeneity at the interface. It is related to the parameter $\alpha_m$ introduced by~\citealt{Monroe2004Effect} as $\alpha_d=1-\alpha_m$ and the parameters $\delta^{\text{etrode}}$  and $\delta^{\text{elyte}}$ used by ~\citealt{McMeeking2019metal}. ~\citealt{diggle1969zinc} found the value of $\alpha_m$ to be 1, giving $\alpha_d=0$. In this work, we have used $\alpha_d=0$ and $\alpha_a=\alpha_c=0.5$. Simulations performed using $\alpha_d=1$ did not show any noticeable difference in the morphological evolution as evident from Fig. S3. The calculation of $\eta_{\text{LC}}$ requires a model for the free energy of the liquid crystal. A liquid crystalline material in the electrolyte introduces an additional grand free energy given by 
\begin{eqnarray}\label{eq:energy}
\Omega_{\text{LC}}[\xi]=\int \text{d}V{\left[\frac{1}{2}K(\nabla \rtvec{n})^2 + \chi (\rtvec{n}\cdot \nabla \xi)^2\right] [1-h(\xi)] }
\end{eqnarray}
Here $\chi$ is the anchoring energy factor accounting for the strong anchoring for the liquid crystal against the surface of the anode. The factor $1-h(\xi)$ ensures that the liquid crystal free energy is non-zero only in the electrolyte phase. The second term is active only at the interface due to the presence of $\nabla \xi$. Using the fact that $|\nabla \xi| \sim 1/\delta $ at the interface and comparing the integral of the second term in Eq. (\ref{eq:energy}) with Eq. (\ref{eq:anch}), we obtain $\chi \sim W\delta/2$. This can be used to obtain estimates of $\chi$ since measured values for $W$ for different liquid crystals are available in literature~\cite{Demus1998handbook}. The additional overpotential due to the liquid crystal can be calculated from its grand free energy contribution using (details in Supporting Information)
\begin{eqnarray}\label{eq:overp}
\frac{n\mathcal{F}}{V_{\ce{M^{z+}}}}\eta_{\text{LC}}=\frac{\delta \Omega_{LC}}{\delta \xi}=\frac{\partial \Omega_{LC}}{\partial \xi} - \frac{\partial \Omega_{LC}}{\partial \nabla \xi}
\end{eqnarray}
where $V_{\ce{M^{z+}}}$ is the molar volume of the metal ion in the liquid crystalline electrolyte. From this equation, we note that the molar volume accounts for the mole fraction share of energy of metal ion out of the total liquid crystal energy. Since liquid crystal relaxation is expected to occur at a time scale much lower than the time scale of diffusion and electrodeposition, we assume equilibrium of the liquid crystal director field in the electrolyte i.e. $\delta \Omega_{LC}/\delta \rtvec{n}=0$. The constraint $\rtvec{n}\cdot \rtvec{n}=1$ on the director field is enforced by the hard constraint method utilizing the Lagrange multiplier technique~\cite{Schwen2017rapid}. Together with evolution equation for the phase-field variable Eq. (\ref{eq:evolxi}), the equations for the evolution of the chemical potential of Li and spatial distribution of the electric potential are also solved and are given in the Supporting Information.

\section{Results and discussion}
\subsection{Electrodeposition morphology}

A $20$ $\mu$m thick Li electrode with perfectly flat interface is used as the initial configuration for the anode upon which Li is electrodeposited. The low initial thickness ensures that the cell has a high energy density due to the high fraction of Li passed per cycle~\cite{zhu2019design, albertus2018status}. The simulation parameters and the details of the initial and boundary conditions are provided in the Supporting Information. We use three metrics to quantify the dendrite growth or suppression during the morphological evolution of the electrode. The first metric, roughness factor is a measure of the unevenness of the Li electrode surface during electrodeposition. A perfectly smooth surface will have a value of zero whereas a surface with dendritic growth will have a high value of roughness factor. The roughness factor RF measures the range of the coordinate profile of the interface. For Fig. \ref{fig:schematic}, the roughness factor is $\text{RF}=x_{max}-x_{min}$. This is one of the definitions of arithmetic average roughness of a surface~\cite{huo2001anomalous}. The second metric is the time required to cause short circuit~\cite{Rosso2001onset} at a given x-coordinate in the 2D mesh. For a given x-coordinate, the short circuit time $t_{\text{sc}}(x)$ is defined as the time when the metal electrode surface reaches that x-coordinate. This gives an indication of the time to short circuit the battery if the counter-electrode was located at that coordinate. The third metric used is related to the arc length of the electrode/electrolyte interface. When the deposition is uneven, the arc length of the interface treated as a curve in two dimensions increases. We measure this deviation using the arc length ratio parameter $\tilde{L}=L/L_0$ where $L$ is the length of the interface at a given time calculated by using the arc length formula for a curve and $L_0$ is the initial length of the interface ($=200$ $\mu$m in our simulations).
Besides quantifying the deviation from an ideal interface, the arc length ratio is also related to the amount of Li consumed at the non-ideal interface resulting in lowering of Coulombic efficiency. In the calculation of these metrics, we used the contour line $\xi=0.5$ as the interface between the two phases.

\begin{figure}[htbp]
    \centering
    \includegraphics[scale=0.6]{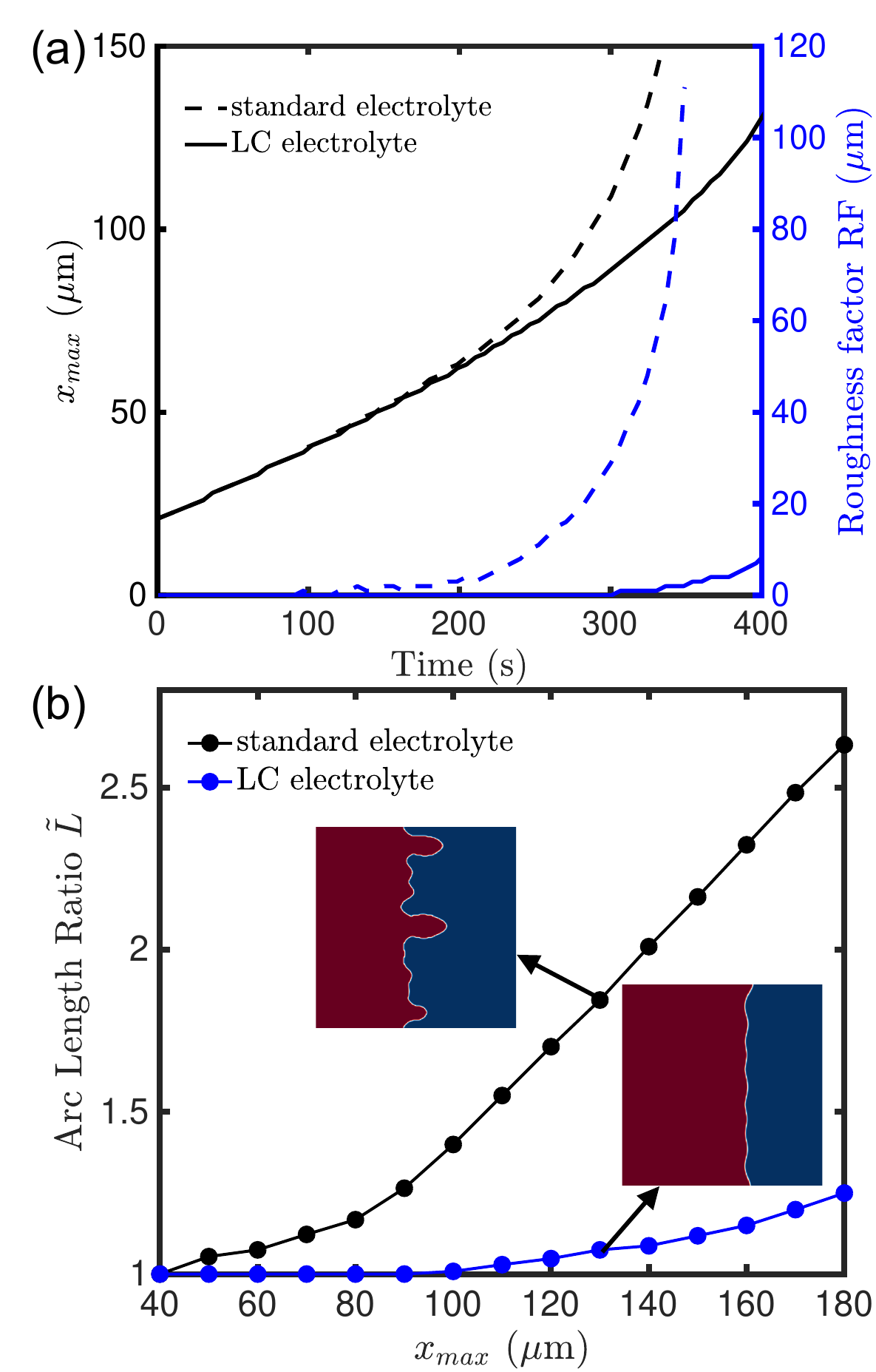}
    \caption{ Metal electrode surface growth with a standard and an LC electrolyte. The LC electrolyte has non-dimensional values of elastic constant $\tilde{K}=39.3$ and anchoring strength $\tilde{W}=20$. (a) Evolution of the metal electrode surface over time measured using the maximum x-coordinate of the interface. Metal electrode surface roughness (in blue) is quantified using the roughness factor $RF=x_{\text{max}}-x_{\text{min}}$ of the interface. (b) The interface arc length ratio measured as the ratio of the length of the interface to the length of the ideal interface is plotted versus $x_{\text{max}}$. The arc length ratio increases rapidly due to the development of dendritic peaks with a standard electrolyte which are suppressed in an LC electrolyte. The insets show the morphology of the metal surface at $x_{\text{max}}=130~\mu$m.}
    \label{fig:roughness_arc}
\end{figure}

We simulated electrodeposition on Li metal anode for the two cases of a conventional liquid electrolyte using the properties of 1 M \ce{LiPF6} in EC:DMC (1:1 volume ratio) solution (hereafter referred to as the standard electrolyte) and a liquid crystalline electrolyte (hereafter referred to as the LC electrolyte). The properties of the electrolytes are given in Supporting Information. For comparison, we assume the dimensionless values of the elastic constant and the anchoring strength to be $\tilde{K}=2\tilde{\delta}^2 \tilde{R}\tilde{T}/\tilde{V}_{\text{Li}^+}=39.3$ and $\tilde{W}=20$. Although LC electrolytes may be engineered to have anisotropic diffusivity~\cite{Tan2016structured}, we use isotropic diffusivity here for the sake of simplicity and comparison with a standard electrolyte. Fig. \ref{fig:roughness_arc}a shows the variation of maximum x-coordinate of the metal electrode surface and the roughness factor as a function of time (black and blue lines respectively). With a standard electrolyte, the metal electrode surface initially grows at a constant velocity with zero roughness factor. The surface starts to roughen at $t\sim 100$ s or when 20 $\mu$m Li has been deposited ($x\sim 40$ $\mu$m at the interface), and the growth rate of the metal starts increasing due to high electric field and \ce{Li+} concentration at the tip of the dendrites~\cite{Hong2018phase}. In contrast, for the case of LC electrolyte, the surface remains uniform even till $t\sim300$ s or when 90 $\mu$m of Li has been deposited.

We observe that once the metal surface begins to roughen and the dendrites start to grow (Fig. \ref{fig:roughness_arc}a), the growth rate increases rapidly since the deposition gets focused at the dendrite tips. Therefore, after the onset of dendritic growth, the maximum x-coordinate of the metal electrode surface at a given time is different in the case of a standard electrolyte and an LC electrolyte. To compare the other two metrics, we used their values when the metal electrode surface attains the same maximum x-coordinate rather than at the same time since this is directly related to the total current passing through the cell. Fig. \ref{fig:roughness_arc}b shows the variation of the arc length ratio $\tilde{L}$ as a function of the maximum x-coordinate of the interface. With a standard electrolyte, the interface arc length ratio quickly becomes greater than one as the metal electrode surface reaches $40$ $\mu$m due to growth of surface perturbations. For $40$ $\mu$m $\lesssim x\lesssim 80$ $\mu$m, several perturbations are generated at the metal electrode surface leading to a positive slope of the $\tilde{L}$ vs $x$ plot. As the metal electrode surface reaches $80$ $\mu$m (60 $\mu$m of Li deposition), these perturbations lead to the growth of three dendrites (movie S1). This leads to an increased slope of the $\tilde{L}$ vs $x_{\text{max}}$ curve. For the case of a LC electrolyte, the interface arc length remains close to the initial value till $x\sim 100$ $\mu$m or for 80 $\mu$m of electrodeposition. After this point, there is a small increase in the arc length ratio due to generation of small surface perturbations, however, none of the perturbations are observed to grow into large dendrites (movie S2).

The discussion above clearly provides a compelling demonstration of the dendrite suppressing nature of LC electrolytes. The liquid crystal is able to suppress the perturbations at the surface through a delicate interplay between bulk and interfacial distortion free energy, oxidation and reduction processes. In contrast to the standard electrolyte, the high current density hotspots during electrodeposition are rarely observed with an LC electrolyte at the interface. The existence of sharp peaks or valleys on the surface will result in sudden change in the orientation of the director field due to the strong anchoring boundary condition at the interface. This will result in an unfavorable high energy configuration of the director field. Our results demonstrate that it is possible to suppress dendrite growth and enhance the fraction of Li passed during cycling using an LC electrolyte. 

\begin{figure*}[htbp]
    \centering
    \includegraphics[scale=0.5]{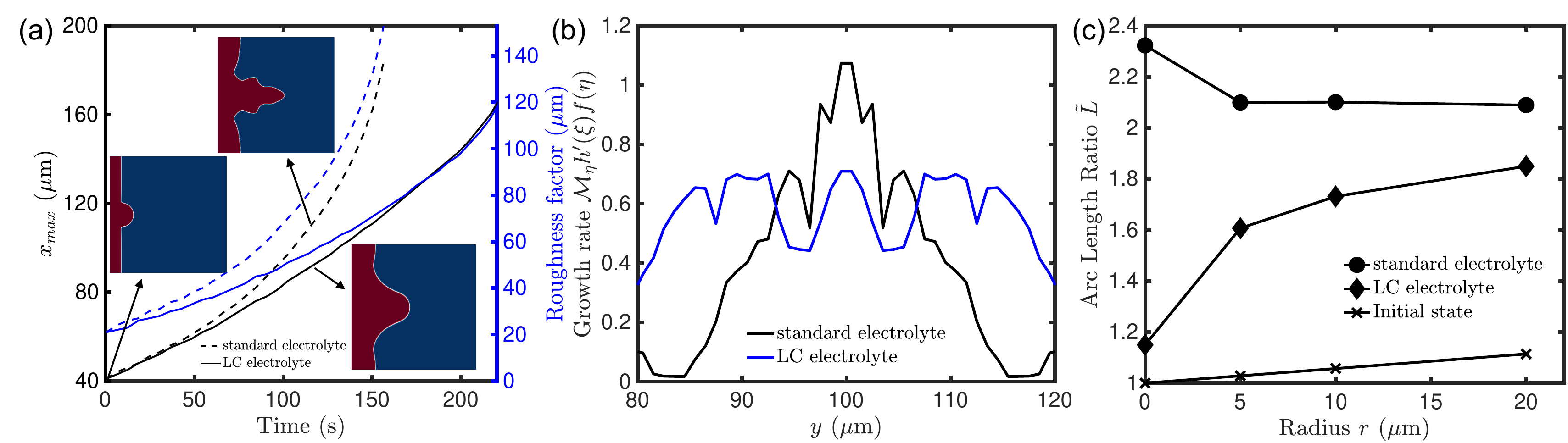}
    \caption{Evolution of a hemispherical nucleus at the metal surface with a standard and an LC electrolyte. The LC electrolyte has values of elastic constant $\tilde{K}=39.3$ and anchoring strength $\tilde{W}=20$. 
    (a) Evolution of maximum x-coordinate (black line) and roughness factor (blue line) of the metal electrode surface for $r=20~\mu$m. The growth of the metal electrode surface with LC electrolyte (solid line) remains approximately linear while the growth with a standard electrolyte (dashed line) is much faster. The insets show the initial condition and morphology after 120 s. (b) Maximum value of growth rate  at a y-coordinate for a standard and LC electrolyte after 31 s for $r=20~\mu$m. The standard electrolyte has highly localized peaks of current density compared to the LC electrolyte. (c) Comparison of arc length ratio for different initial nucleus radii $r$ at the metal surface at $x_{\text{max}}=160$ $\mu$m. The LC electrolyte is more effective at suppressing dendrites for smaller initial surface perturbations.}
    \label{fig:nucl}
\end{figure*}

\textit{Nucleation}-  To understand the rearrangement of the director field at a rough surface and its effects on electrodeposition, we generate an initial perturbation on the the metal electrode surface and simulate electrodeposition under an applied overpotential. The perturbation is a hemispherical nucleus with three different radii: $5~\mu$m, $10~\mu$m and 20 $\mu$m generated by setting the initial condition for the phase-field variable $\xi=1$ inside the hemisphere. The director field of the liquid crystal reorients in response to the perturbation of the metal electrode surface (Fig. S2). Due to anchoring energy of the LC electrolyte, the director field becomes tangent to the metal electrode surface at the interface. The director field in the vicinity of the interface also changes to minimize the distortion free energy which is proportional to the bulk elastic constant. Fig. \ref{fig:nucl}a shows the maximum x-coordinate and roughness of the metal surface with time for the standard and LC electrolyte. The insets show the initial condition and the metal surface after 120 s of electrodeposition. Electrodeposition with the standard electrolyte leads to the development of sharp peaks from the initial nucleus. These peaks originate from the high current density hotspots and encourage faster electrodeposition by attracting metal ions due to the high electric fields generated, as explained in Ref. ~\citenum{Hong2018phase}. In contrast, the LC electrolyte prevents the formation of sharp peaks at the interface due to the anchoring energy, leading to an approximately constant growth velocity and a spatially homogeneous growth at the metal surface. The function $\mathcal{M}_{\eta} h'(\xi) f(\eta)$, which is the growth rate of the metal surface due to the overpotential, is plotted for the standard and LC electrolyte  in Fig. S4 at $t=31$ s. The standard electrolyte case has a more localized brown region at the tip of the hemisphere, showing a high current density hotspot as compared to the LC electrolyte case.  This point is elucidated in Fig. \ref{fig:nucl}b showing the maximum of the metal surface growth rate due to overpotential, $\mathcal{M}_{\eta}h'(\xi)f(\eta)$ at each y-coordinate in the 2D domain. The maximum in the growth rate for a given y-coordinate occurs at the interface where the electrodeposition reaction occurs. The growth rate for the standard electrolyte is much more localized at the tip of the nucleus ($y=100~\mu$m) compared to that for an LC electrolyte.  The arc length ratio plotted as a function of the maximum x-coordinate of the metal surface for different initial nucleus radii also increases faster for the standard electrolyte compared to the LC electrolyte (Fig. S7). Fig. \ref{fig:nucl}c shows the variation of the arc length ratio at $x_{\text{max}}=160$ $\mu$m with the radius of the nucleus. The arc length ratio obtained using an LC electrolyte decreases as the size of the initial perturbation decreases while it remains almost constant with a standard electrolyte. This variation can be used to determine the initial roughness of the metal anode sample to design for a given thickness of electrodeposited metal and final roughness/arc length ratio that can be tolerated.

\subsection{Liquid crystalline electrolyte selection metrics}
As outlined earlier, for use in a battery, the liquid crystalline electrolyte must satisfy several properties in addition to the metrics we identified in the previous section.  We begin by exploring the oxidative stability of liquid crystalline molecules and then explore the metrics related to electrodeposition.

\textbf{Oxidative Stability:} One necessary but not sufficient condition~\cite{Peljo2018electrochemical} is that the highest occupied molecular orbital (HOMO) level of the electrolyte should lie below the electrochemical potential of the cathode $\mu_C$. A large negative HOMO level (referenced to vacuum level) would ensure sufficient stability of the electrolyte and prevent decomposition. We obtained the HOMO level of the electrolyte using DFT calculations of neutral and positively charged liquid crystal molecules which have been well benchmarked against experimental data~\cite{Pande2019descriptors}. As shown in Table \ref{tab:adsdft}, we find that liquid crystalline electrolyte molecules considered have moderate cathode stability (HOMO level $\sim -7$ eV) relative to conventional small organic molecules used (ethylene carbonate, dimethyl ether) in traditional liquid electrolytes, that have HOMO levels between $-8$ and $-10$ eV.   Among the liquid crystal molecules considered, 5CB is the most stable due to the cyano-group, while the addition of an ether functionalization reduces the HOMO level a bit in 5COB.  We find the other ether and ester-based liquid crystalline molecules (CID: 170879 and 2058335) have similar stability while among the molecules considered, MBBA is the least stable due to the presence of the imine functional group.  As shown here, the HOMO levels of these molecules can be manipulated by lowering the acceptor number of the end functional groups~\cite{Pande2019descriptors}, such as those demonstrated by carbonate-based (stable upto 2 V vs. Li/\ce{Li+})~\cite{Kato2015liquid} or cyano-based liquid crystals (stable upto 3 V vs. Li/\ce{Li+})~\cite{Champagne2019liquid}.  Engineering the cathode stability will be an important challenge to realize these liquid crystalline electrolytes in a practical system.

\begin{figure}[htbp]
    \centering
    \includegraphics[scale=0.75]{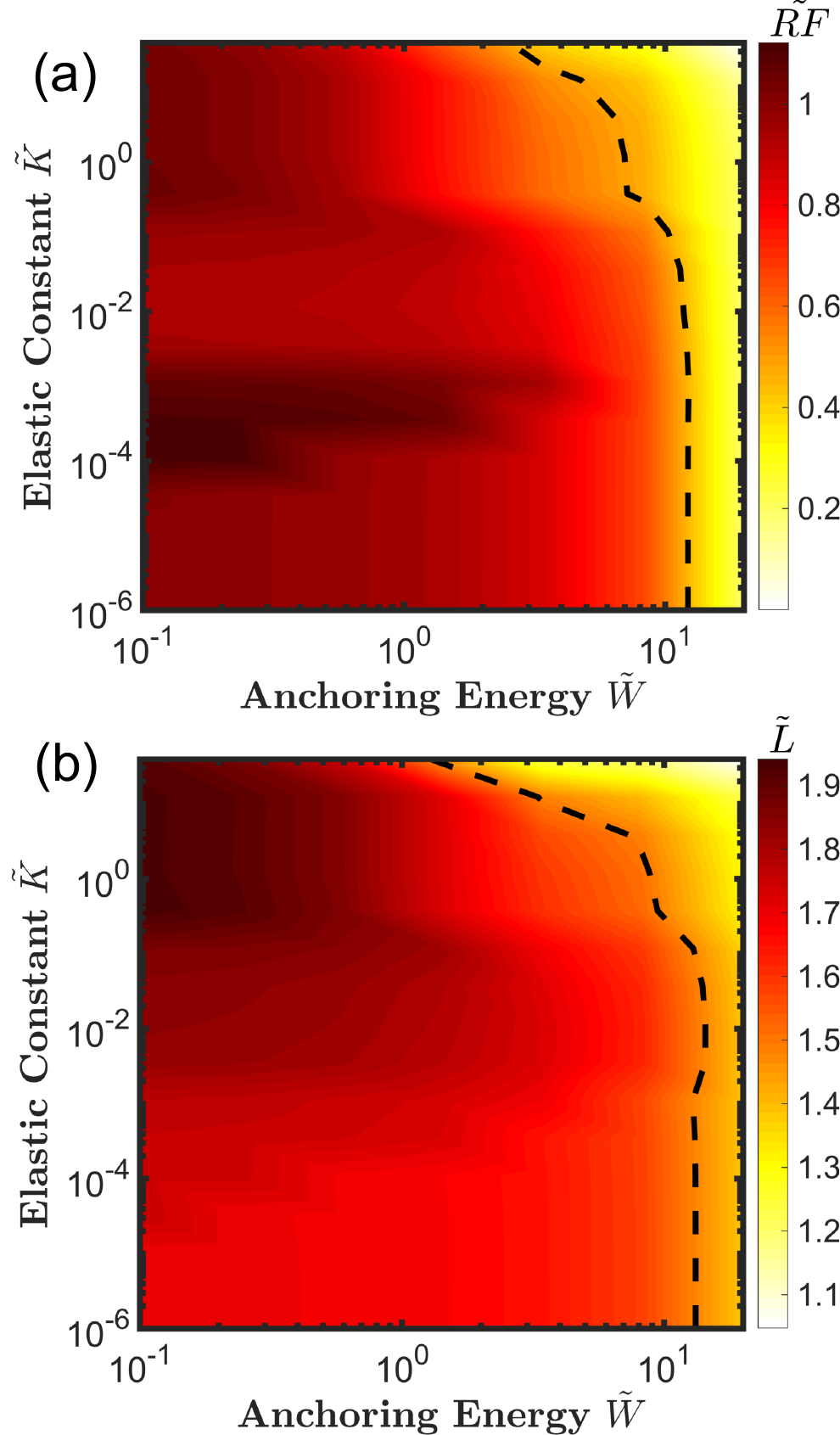}
    \caption{\label{fig:contourall} Effect of elastic constant $K$ and anchoring strength $W$ on the dendrite suppression metrics: (a) roughness factor $\tilde{RF}=RF/RF_0$ measured at $x_{\text{max}}=90$ $\mu$m where $RF_0$ is the roughness using the standard electrolyte (b) arc length ratio $\tilde{L}$ measured at $x_{\text{max}}=120~\mu$m. The anchoring strength affects the dendrite suppression capability of the LC electrolyte much more than the elastic constant. The contour lines marks $\tilde{RF}=0.5$ and arc length ratio=1.5.}
\end{figure}

\textbf{Dendrite Suppression}: Having explored the oxidative stability, we will now understand the effect of the properties of liquid crystal on the metrics for dendrite suppression.   We performed phase-field simulations using a range of values of these parameters, i.e. elastic constant $K$ and anchoring strength $W$ to construct a two dimensional phase-diagram. Fig. \ref{fig:contourall} shows the variation of two metrics, roughness factor $\tilde{RF}$ and interface arc length ratio $\tilde{L}$ with the values of these parameters. The variation of short circuit time $\tilde{t}_{sc}$ is presented in Fig. S8. The roughness factor, arc length ratio and short circuit time are calculated when the metal electrode surface reaches $x=90, 120$ and $150$ $\mu$m respectively. The roughness factor is represented in non-dimensional form as a ratio of its value obtained using LC electrolyte to the that obtained with a standard electrolyte. For all metrics, we observe that the capability to suppress dendrites is improved as the elastic constant and the anchoring strength of the LC electrolyte is increased. From the plots, we observe that the anchoring strength $W$ is more influential for dendrite suppression than the bulk elastic constant. The arc length ratio value of 1.5 is plotted as a contour line in Fig. \ref{fig:contourall}b. The region to the right of the line has better dendrite suppression capability. A higher elastic constant further increases the region of stability.


The anchoring energy with lithium metal presents a new opportunity for the development of dendrite-suppressing LC electrolytes. There are two possible directions: development of new high anchoring liquid crystal materials and engineering surfaces using, for example, coupling agents to obtain a higher degree of coupling between the metal anode and the liquid crystal resulting in high anchoring energy~\cite{kahn1973orientation}. There is limited experimental data on the anchoring energy and the influence of interfaces on detailed positional order of liquid crystals since it requires a technique sensitive to the position of molecules~\cite{Jerome1991surface}. Further, anchoring of a liquid crystal with a surface involves a variety of intermolecular interactions and is hard to model using first-principles directly without simulating thousands of atoms~\cite{Yokoyama1988surface}. 

\begin{figure}
    \centering
    \includegraphics[scale=0.7]{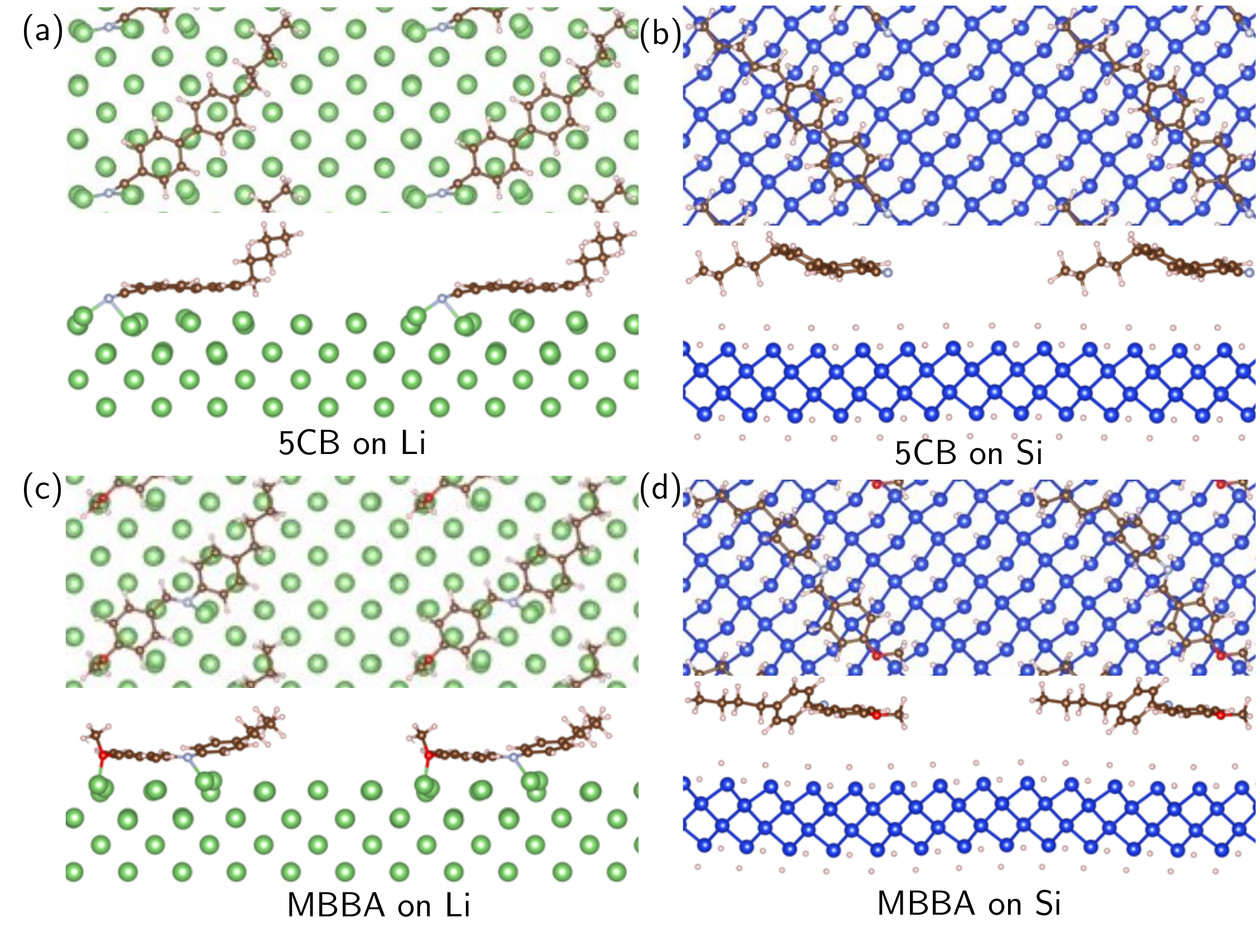}
    \caption{Final geometry (top and front views) obtained on DFT relaxation of liquid crystal molecules (a) 5CB on Li, (b) 5CB on Si, (c) MBBA on Li, (d) MBBA on Si surface. The values of adsorption energies are given in Table \ref{tab:adsdft}. Compared to Si surface, the Li surface favors strong bonding resulting in higher adsorption energy and stronger anchoring of the liquid crystal molecules\label{fig:anchad}.}
\end{figure}

To obtain a theoretical estimate for the anchoring strengths of liquid crystals with Li metal, we performed DFT calculations of different liquid crystal molecules on Li (100) surface using adsorption energy as the descriptor. This is motivated by scanning tunneling microscopy experiments showing the presence of a liquid crystal surface monolayer which is immobilized in contact with a solid whose strength is controlled by physisorption~\cite{Foster1988imaging,Spong1989imaging}. The anchoring of liquid crystal is determined by the minimization of the interfacial energy $\gamma_s$ in the presence of the liquid crystal molecules~\cite{Jerome1991surface}
\begin{equation}
    \gamma_s(\rtvec{n}) = \gamma_s^{\text{surf}}(\rtvec{n}) + \gamma_s^{\text{ad}}(\rtvec{n},\rtvec{a_{\text{ad}}}) 
\end{equation}
where $\gamma_s^{\text{surf}}(\rtvec{n})$ gives the interfacial energy without adsorption and $\gamma_s^{\text{ad}}(\rtvec{n},\rtvec{a_{\text{ad}}})$ accounts for the adsorption energy of the molecule dependent on the molecular orientation $\rtvec{a}_{\text{ad}}$.
The molecules studied contain different functional groups representing the diversity of commonly studied liquid crystals. We also performed the same calculation for 5CB and MBBA liquid crystal on silicon surface to calibrate our results with those obtained from molecular dynamics simulations~\cite{Pizzirusso2012predicting}. Our calculations substantiate the role of adsorption in controlling the molecular alignment close to the interface (Fig. \ref{fig:anchad}). 

We find four liquid crystal molecules that have adsorption energies on Li surface much higher than that on Si surface (Table \ref{tab:adsdft}). Therefore, we expect the director field of the liquid crystal at the interface to be more strongly anchored by the Li surface compared to the Si surface due to a higher $\gamma_s^{\text{ad}}$.  Our simulations assume a planar anchoring of the liquid crystals but a tilted anchoring may also be used to suppress dendrites. A homeotropic anchoring, however, cannot alter the kinetics of electrodeposition since it can accommodate uneven morphology of the metal anode without an increase in bulk distortion or anchoring free energy~\cite{Berreman1972solid}.

It is also worth noting that liquid crystalline properties can be changed by temperature~\cite{degennes1974liquid} and engineering the density of packing or particle shapes~\cite{vanAnders2014shape}. These can lead to the emergence of directional entropic forces that align neighboring particles. ~\citealt{Weng2015anchoring} reported an increase in anchoring energy by using vertical alignment and polymerized surfaces generated by ultraviolet irradiation-induced phase separation. Further, nanopatterning of the surfaces using surface lithography can lead to generation of nanogrooves which can be used to tune the anchoring strength~\cite{Lee2001alignment,Gear2015engineered}. The dendrite suppression metrics can be further improved by increasing the molar volume of Li in the LC electrolyte (Fig. S9). This can be achieved, in practice, by tuning the ion-solvent interactions~\cite{marcus-ionvolumes}.

\begin{table}[htbp]
  \centering
  \caption{\label{tab:adsdft}Adsorption energy (normalized by surface area) of liquid crystal molecules on Li and Si surface from DFT simulations. Also shown are the HOMO levels of the molecules relative to vacuum and their transition temperatures $T_{\text{NI}}$ obtained from Refs.~~\citenum{Pestov2003} and~\citenum{thiem1992transition}.}
    \begin{tabular}{llclr}
    \hline\hline
    PubChem CID & Common name  & $T_{\text{NI}}$ ($^{\circ}$C) & Adsorption Energy (J/m$^2$) & HOMO level (eV) \\
    \hline
     92319 & 5CB  & 35.3 & \begin{tabular}{@{}c@{}}-0.066 (Li)\\ -0.009 (Si)\\ -0.033 (Fluorinated Li)  
     \end{tabular} & -7.69 \\
     33363 & MBBA  & 48.0 &  \begin{tabular}{@{}c@{}}-0.079 (Li)\\ -0.018 (Si)\end{tabular} & -6.79 \\
     104171 & 5COB & 68.0 &  -0.007 (Li) & -7.36 \\
     170879 & -  & 15.0 &  -0.040 (Li) & -7.38 \\
     2058335 & -   & - &  -0.045 (Li) & -7.47\\
          \hline
     \hline
\end{tabular}
\end{table}


\section{Concluding Remarks}
We have suggested a new material class, liquid crystals, as candidate electrolytes for Li metal anodes.  Additional energetic contributions due to anchoring and distortion provide a new approach to suppress the onset of dendrite formation. We implemented a phase-field model which accounts for the overpotential due to the anchoring and bulk distortion of the liquid crystal, and combine them with DFT calculations to explore the design space for candidate liquid crystalline electrolytes. Using the phase-field model, we demonstrate smooth electrodeposition and spontaneous dendrite suppression with liquid crystalline electrolytes. To our knowledge, this work presents the first demonstration of dendrite suppressing character of liquid crystalline electrolytes. We identify the anchoring strength as an important tuning property that determines the degree of dendrite suppression as suggested by the values of the metrics. Having identified the design rules, we develop descriptors to propose material formulations that can satisfy these criteria.  We establish a set of descriptors to screen for candidate liquid crystalline electrolytes: surface adsorption energy to analyze anchoring strength and other metrics for use in a battery such as order-disorder transition temperature, HOMO level for oxidative stability. Given the very large design space of liquid electrolytes, it is undoubtedly possible to break these design trade-offs through electrolyte engineering.

\begin{acknowledgement}
The authors thank Y.-M. Chiang, B.A. Helms, P.D. Frischmann, V. Pande and D. Krishnamurthy for helpful discussions. Z.A., Z.H. and V.V. acknowledge support from the Advanced Research Projects Agency Energy (ARPA-E) under Grant DE-AR0000774.
\end{acknowledgement}

\begin{suppinfo}

Details of MOOSE simulation parameters, initial and boundary conditions, derivation of overpotential and other phase-field equations, evolution of metal surface, dendrite suppression metrics and movies for Li surface evolution. The code is made freely available from https://github.com/ahzeeshan/electrodep.

\end{suppinfo}

\bibliography{refs}


\section*{Supplementary material}

# Supporting Information for

# Design rules for liquid crystalline electrolytes for enabling dendrite-free lithium metal batteries

Zeeshan Ahmad,[†] Zijian Hong,[†] and Venkatasubramanian Viswanathan[*,†,‡]

†Department of Mechanical Engineering, Carnegie Mellon University, Pittsburgh, Pennsylvania 15213, USA

‡Department of Chemical Engineering, Carnegie Mellon University, Pittsburgh, Pennsylvania 15213, USA

E-mail: venkvis@cmu.edu

# Interface between Li metal and solid electrolyte

# MOOSE simulation parameters

All simulations were performed on a two-dimensional mesh 200 $\mu$m by 200 $\mu$m in size sampled by 200 by 200 grid points using an adaptive time step of maximum 0.01 s. Numerical integration was performed using the `bdf2` scheme while the system of PDEs was solved using Newton method and the single matrix preconditioner as implemented in MOOSE.[1] An overpotential of 200 mV was used for all simulations.

### Initial Conditions

The metal electrode is located in the region $0 \leq x \leq 20$ $\mu$m ($\xi = 1$) and the electrolyte in the region 20 $\mu$m $\leq x \leq$ 200 $\mu$m ($\xi = 0$) with the gradual variation represented by a `tanh` function. The Li mole fraction in the electrolyte is set based on 1 M Li$^+$ concentration in a carbonate based electrolyte. The director field of the liquid crystal **n** is set to $(0,1)^T$ throughout the electrolyte. The potential is set to -0.2 V on the electrode side and 0 V on the electrolyte side. For simulations with an initial perturbation, a semicircular region at $y = 100$ $\mu$m initially has $\xi = 1$ with the required radius.

### Boundary Conditions

In general, Dirichlet boundary conditions are employed for the left and boundaries and Neumann boundary conditions for the top and bottom boundaries. The phase-field variable is set to 1 on the left boundary and 0 on the right boundary The potential is set to -0.2 V in the left boundary and 0 V at the right boundary. The derivatives of phase-field variable and chemical potential are set to zero at the top and bottom boundaries. This can be ensured through, for example, strong anchoring at the interfaces or generation of grooves perpendicular to the plane of the simulation.[2] The boundary condition for the director is

$\mathbf{n} = (0, 1)^T$ at the right boundary and $\mathbf{n}$ is parallel to the tangent to the surface of the metal electrode at the interface. The latter boundary condition is imposed using the soft constraint method[3] i.e. by adding the energy contribution due to anchoring to the grand free energy in Eq. (5) in the main text.

## Anchoring Energy

We use the form of anchoring energy proposed by Rapini and Papoular[4,5]

$$F_{\text{anch}} = \frac{1}{2} W [1 - (\mathbf{n} \cdot \mathbf{n}_p)^2] \tag{1}$$

where $\mathbf{n}_p$ is the preferred direction for the liquid crystal director in contact with a surface. This form might be further reduced to $F_{\text{anch}} = -W(\mathbf{n} \cdot \mathbf{n}_p)^2/2$ since the first term is a constant offset to the free energy. In two dimensions, if $\mathbf{t}$ is the preferred direction and $\mathbf{v} \perp \mathbf{t}$, the anchoring free energy may be written as $F_{\text{anch}} = -W(\mathbf{n} \cdot \mathbf{t})^2$ or $F_{\text{anch}} = +W(\mathbf{n} \cdot \mathbf{v})^2/2$ since the two differ only by a constant. The latter form is the one we have used in this work.

## Overpotential expression

The overpotential due to the LC electrolyte in terms of its properties is given by:

$$\frac{n\mathcal{F}}{V_M} \eta_{\text{LC}} = \frac{\delta \Omega_{LC}}{\delta \xi} = \frac{\partial \Omega_{LC}}{\partial \xi} - \frac{\partial \Omega_{LC}}{\partial \nabla \xi} - h'(\xi) \left[ \frac{1}{2} K (\nabla \mathbf{n})^2 + \chi (\mathbf{n} \cdot \nabla \xi)^2 \right]$$
$$+ 2\chi (1 - h(\xi)) \left[ \text{div}(\mathbf{n})(\mathbf{n} \cdot \nabla \xi) + \mathbf{n} \cdot \nabla (\mathbf{n} \cdot \nabla \xi) \right]$$

## Other Phase-field Equations

In the formalism developed by Plapp and Hong et al.,[6,7] the local Li-ion mole fraction can be written in terms of the chemical potential of Li, $\mu$ as:

$$c_{\text{Li}^+} = \frac{\exp\left[(\mu - \epsilon^l)/RT\right]}{1 + \exp\left[(\mu - \epsilon^l)/RT\right]}(1 - h(\xi)) \qquad (2)$$

where $\epsilon^l = \mu^{0l} - \mu^{0N}$ is the difference in the chemical potential of Li and neutral components in the electrolyte phase at the standard state. We use $l$ as the superscript for the electrolyte and $s$ for the electrode. The equation for evolution of chemical potential is:

$$\frac{\partial \mu}{\partial t} = \frac{1}{\chi_\mu}\left[\nabla \cdot \frac{Dc_{\text{Li}^+}}{RT}(\nabla\mu + n\mathcal{F}\nabla\phi) - \frac{\partial h(\xi)}{\partial t}\left(c^s \frac{C_m^s}{C_m^l} - c^l\right)\right] \qquad (3)$$

where $C_m^s$ and $C_m^l$ are the site densities of Li in the electrode and electrolyte phases, $\phi$ is the electric potential, $n$ is the number of electrons transferred and $c$ is the concentration of Li. The susceptibility $\chi_\mu$ is given by

$$\chi_\mu = \frac{\partial c^l}{\partial \mu}[1 - h(\xi)] + \frac{\partial c^s}{\partial \mu}h(\xi)\frac{C_m^s}{C_m^l} \qquad (4)$$

For obtaining the electric potential, we solve the conduction equation:

$$\nabla \cdot (\sigma \nabla \phi) = n\mathcal{F}C_m^s \frac{\partial \xi}{\partial t} \qquad (5)$$

where $\sigma = \sigma^s h(\xi) + \sigma^l(1 - h(\xi))$ is the electronic conductivity.

# Properties of the Electrolytes

1 M LiPF$_6$ in EC/DMC (1:1 volume ratio) is used for obtaining electrolyte properties like diffusivity $D$, conductivity $\sigma$ etc. For the LC electrolyte, the molar volume of Li$^+$ is assumed to be 5.022E-05 mol/m$^3$ [8] in the calculation of overpotential. All properties are tabulated in Table S1.

Table S1: Parameters used in the phase-field model. The code is available from https://github.com/ahzeeshan/electrodep

| Properties | Value in S.I. units | S.I. Units | Normalized Value |
|---|---|---|---|
| $RT$ | 2494.2 | J/mol | 9.870E-01 |
| $\gamma$ [9] | 0.556 | J/m$^2$ | 2.200E-01 |
| $\delta$ | 1.00E-06 | m | 1.000E+00 |
| $\kappa$ | 8.34E-07 | J/m | 3.300E-01 |
| $b$ | 6.67E+06 | J/m$^3$ | 2.640E+00 |
| $D$ [10] | 3.20E-10 | m$^2$/s | 3.197E+02 |
| $\sigma^s$ | 1.0E+07 | S/m=A/(Vm) | 3.953E+12 |
| $\sigma^l$ [10] | 1.19 | S/m=A/(Vm) | 4.704E+05 |
| $\alpha$ [11] | 0.5 | - | 0.5 |
| $i_0$ [12] | 15 | A/m$^2$ | 5.929E+00 |
| $L_\sigma$ [11] | 2.50E-06 | m$^3$/(Js) | 6.318E+00 |
| $L_\eta = \gamma V_{\text{Li}} i_0 / \mathcal{F}\kappa$ [7] | 1.35E-03 | 1/s | 1.347E-03 |
| $\mathcal{F}$ | 9.6485E+04 | C/mol | 3.814E+01 |
| $V_{\text{Li}^+}$ [8] | 5.022E-05 | m$^3$/mol | 5.022E-02 |
| $V_{\text{Li}}$ [8] | 1.30E-05 | m$^3$/mol | 1.300E-02 |
| $\mu^l$ | 173.1960308 | J/mol | 6.854E-02 |
| $\epsilon^l$ | 6565.678657 | J/mol | 2.598E+00 |
| $\mu^s$ | 34458.64644 | J/mol | 1.364E+01 |
| $\epsilon^s$ | -34458.64394 | J/mol | -1.364E+01 |
| $c_0$ | 1.00E+03 | mol/m$^3$ | 1.000E+00 |

# Property Normalization/Scaling

The normalization constants for the different quantities are presented in Table S2.

Table S2: All variables are normalized using the following values in SI units, e.g. lengths are normalized by $10^{-6}$ m.

| Variable | Value |
|---|---|
| Length | 1.00E-06 m |
| Mass | 2.527 kg |
| Time | 1 s |
| Moles | 1.00E-15 mol |
| Charge | 2.53E-12 C |
| Voltage | 1.00E+00 V |

# Metrics for dendrite suppression

For Fig. 1 (in the main text), if the interface is given by the curve $x = f(y)$, the arc length of the interface can be calculated using $\int \mathrm{d}y \sqrt{1 + f'(y)^2}$ which can be converted to a discrete summation.

Table S3: Elastic constants of common liquid crystalline materials. Data obtained from Refs.[13–19]

| Molecule Name | $K_{11}$ (N) | $\tilde{K}_{11}$ (Normalized) |
|---|---|---|
| PAA: p-azoxyanisole | 9.00E-12 | 3.56E-06 |
| MBBA | 7.48E-12 | 2.96E-06 |
| Nematic 5CB: 4-n-pentyl-4-cyano biphenyl | 8.00E-12 | 3.17E-06 |
| Py-7: 5-n-heptyl-2-(4'- cyanophenyl)-pyrimidine. | 5.04E-12 | 1.99E-06 |
| CB-7: 4-n-heptyl-4' -cyanobiphenyl. | 8.55E-12 | 3.38E-06 |
| PCH-7: trans-4-n -heptyl-(4 ' -cyanophenyl)-cyclohexane. | 1.01E-11 | 4.00E-06 |
| CCH -7: trans, trans -4-n -heptyl:'4 ' -cyanobicyclohexane. | 7.24E-12 | 2.86E-06 |
| C -7: trans-4' -cyanophenyl-4-n-heptylcyclohexanoate. | 7.75E-12 | 3.06E-06 |
| C-5: trans4' -cyanophenyl-4-n -pentylcyclohexanoate. | 7.75E-12 | 3.07E-06 |
| 7CB | 7.3E-12 | 2.89E-06 |
| 6CB | 3.80E-12 | 1.50E-06 |

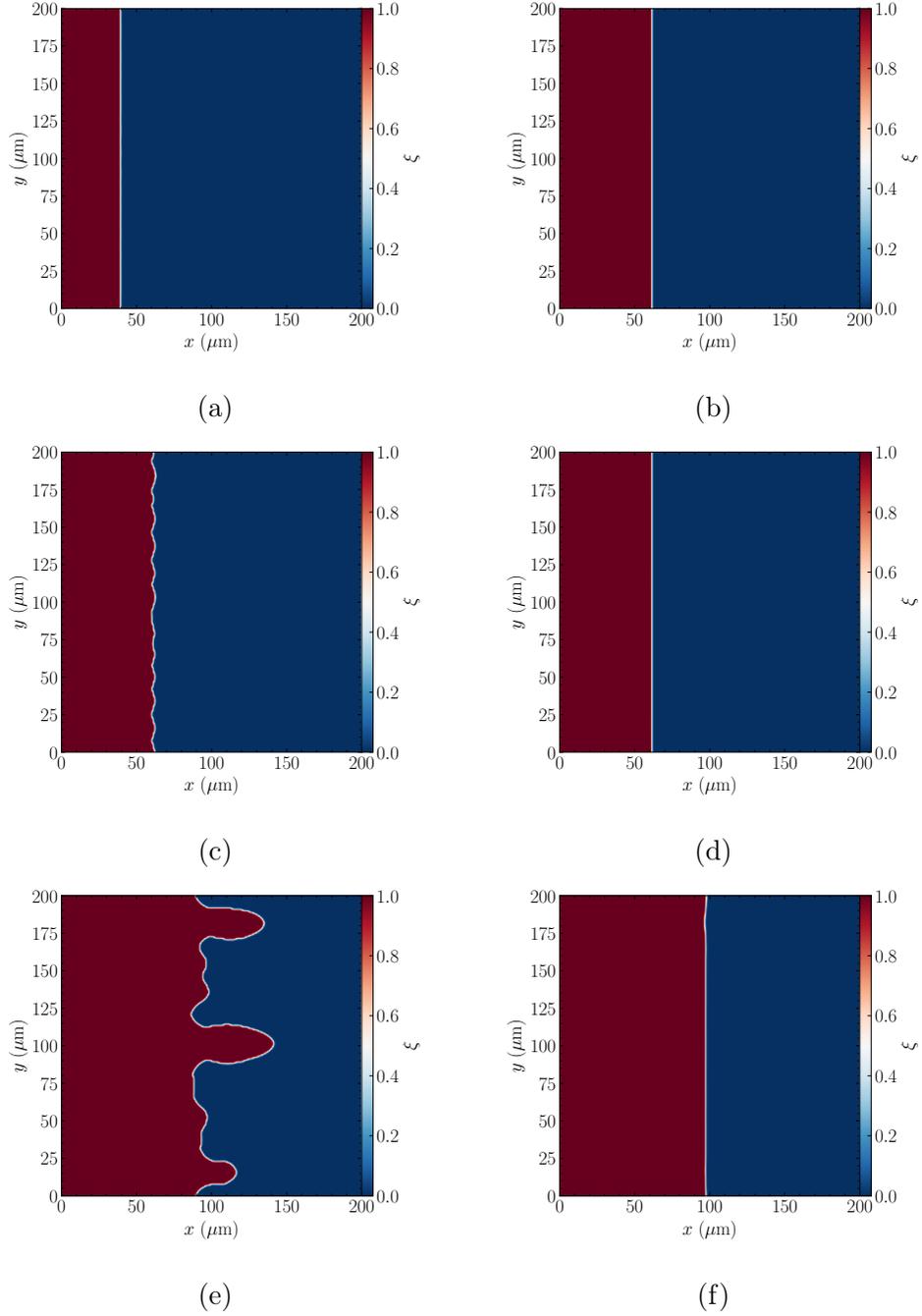

Figure S1: Growth of metal surface with a standard [(a), (c) and (e)] and LC electrolyte [(b), (d) and (f)] at different times: 100 s [(a), (b)], 200 s [(c), (d)], 330 s [(e), (f)]. The dendrites start to nucleate after 200 s and eventually develop into large dendrites after 330 s with the standard electrolyte. In contrast, the interface is quite stable with the LC electrolyte up to 330 s. This clearly indicates that the LC electrolyte can help suppress dendrite formation.

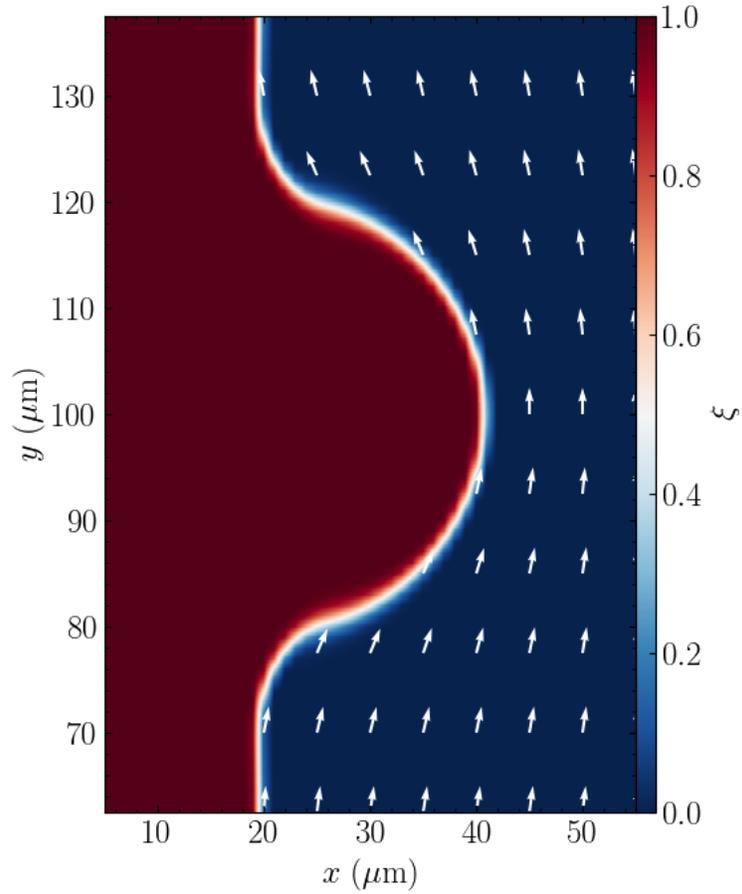

Figure S2: Distortion of the liquid crystal director field **n** due to an initial hemispherical nucleus. The arrows show the spatial distribution of the director field of the liquid crystal in the vicinity of the surface perturbation. It is clearly shown that the director field reorients along the hemispherical nucleus which is caused by the anchoring energy.

8

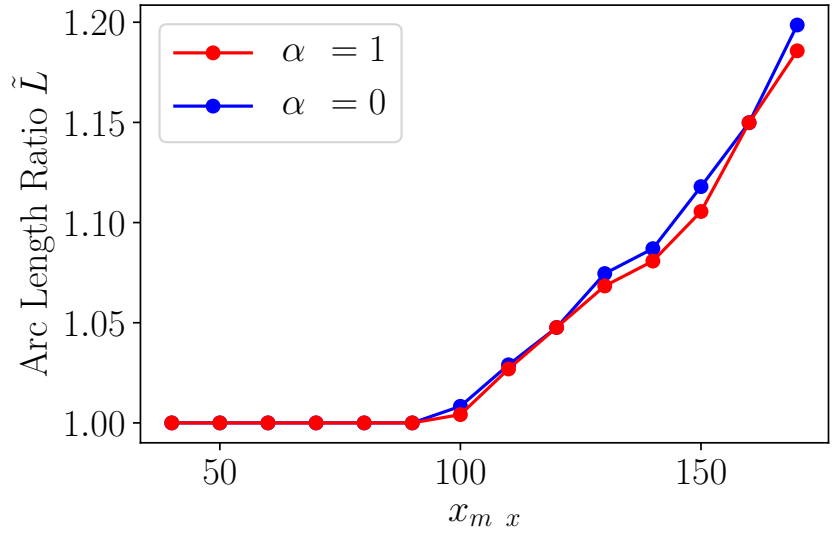

(a)

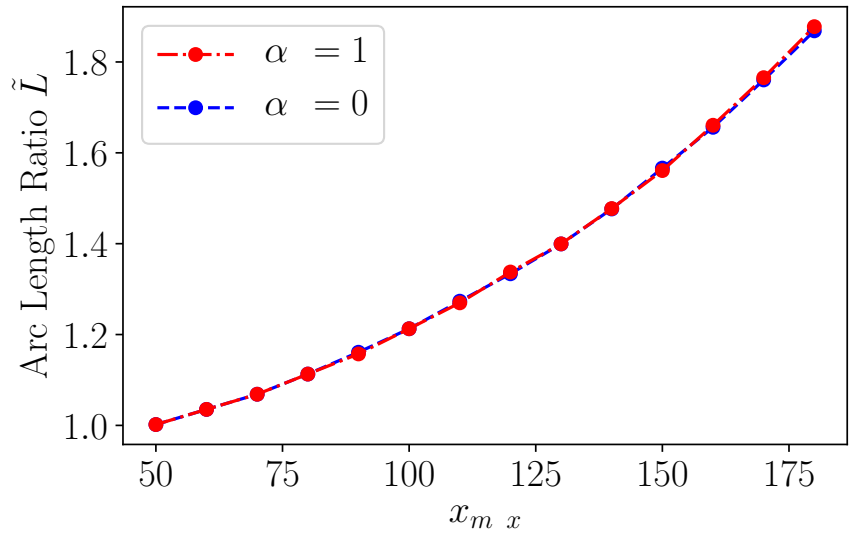

(b)

Figure S3: Comparison of the arc length ratio $\tilde{L}$ for $\alpha_d = 0$ and 1 for the case of an initial condition of (a) smooth metal surface and (b) 40 $\mu$m nucleus at the metal surface (same as presented in the main text).

9

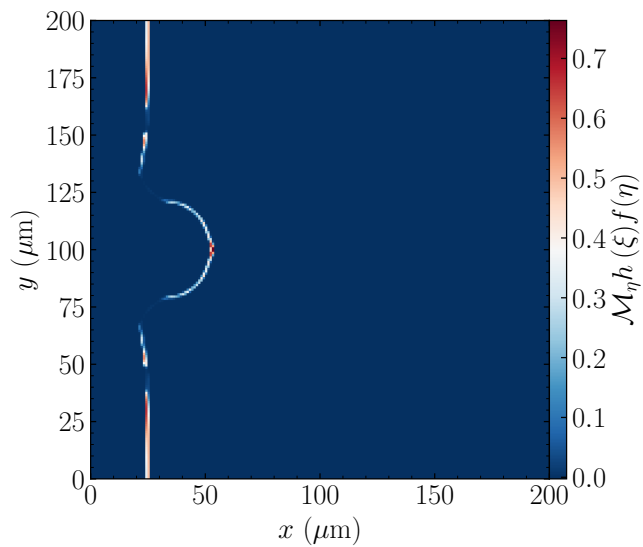

(a)

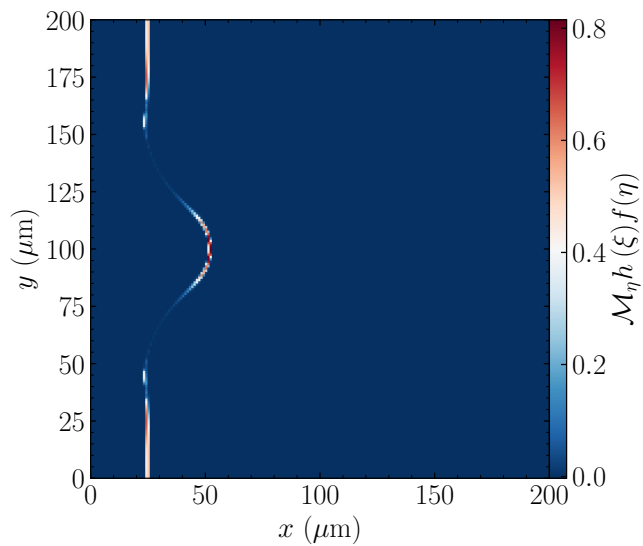

(b)

Figure S4: Comparison of the electrodeposition kinetics for the case of (a) standard electrolyte and (b) LC electrolyte at time=31 s. Higher the value of $\mathcal{M}_\eta h'(\xi)f(\eta)$, the higher electrodeposition rate. The current density is much more localized at the tip of the nucleus in the case of a standard electrolyte compared to the LC electrolyte.

10

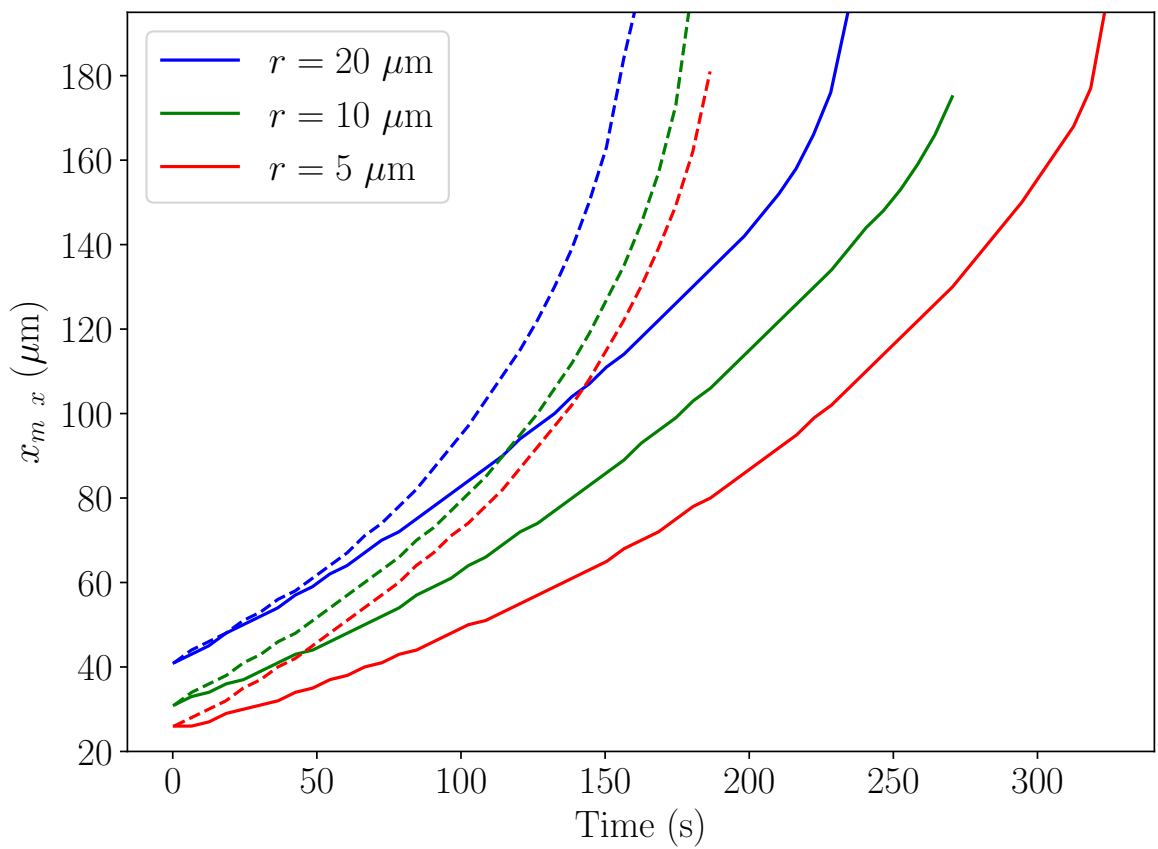

Figure S5: Evolution of the maximum x-coordinate for a standard electrolyte (dashed line) and LC electrolyte (solid line) with time. The LC electrolyte has a slower growth.

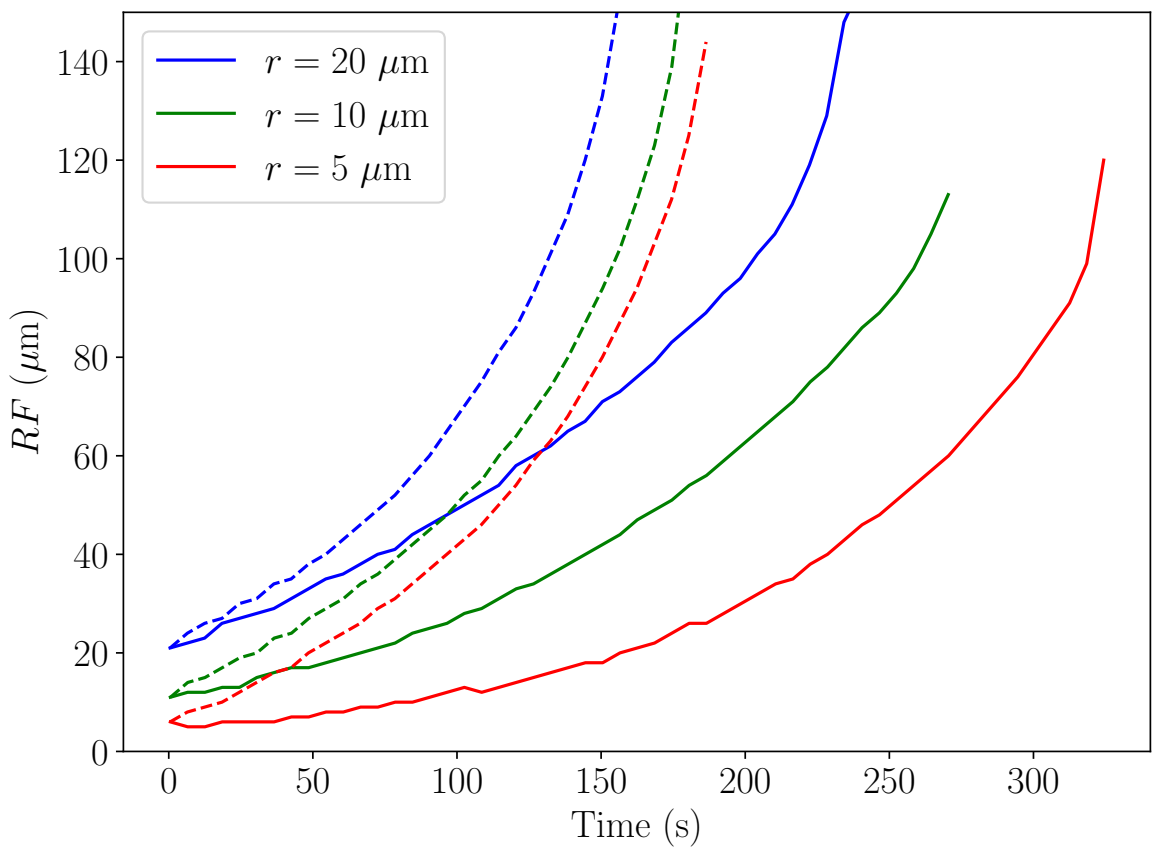

Figure S6: Evolution of the roughness factor $RF$ for a standard electrolyte (dashed line) and LC electrolyte (solid line) with time showing the same trend as $x_{\max}$.

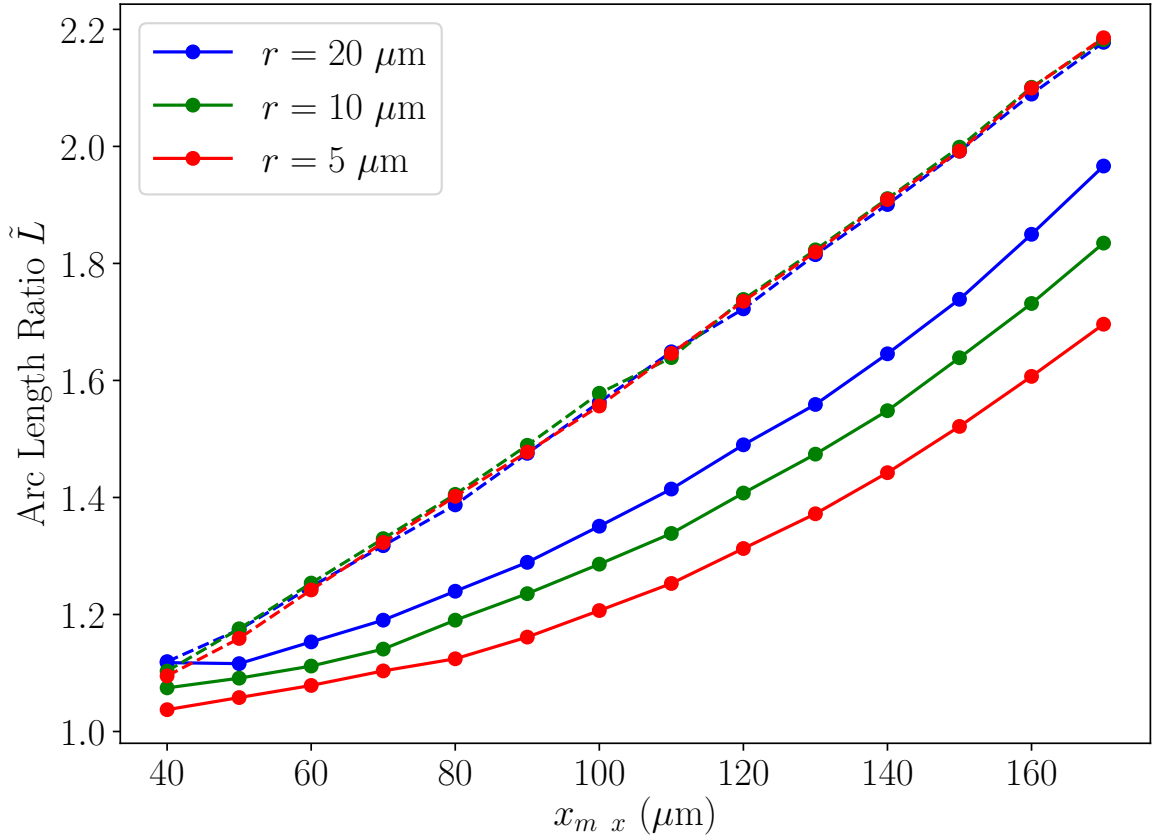

Figure S7: Evolution of the interface arc length ratio with maximum x-coordinate for a standard electrolyte (dashed line) and LC electrolyte (solid line). The LC electrolyte has a slower growth of the arc length ratio due to suppression of sharp dendritic peaks.

13

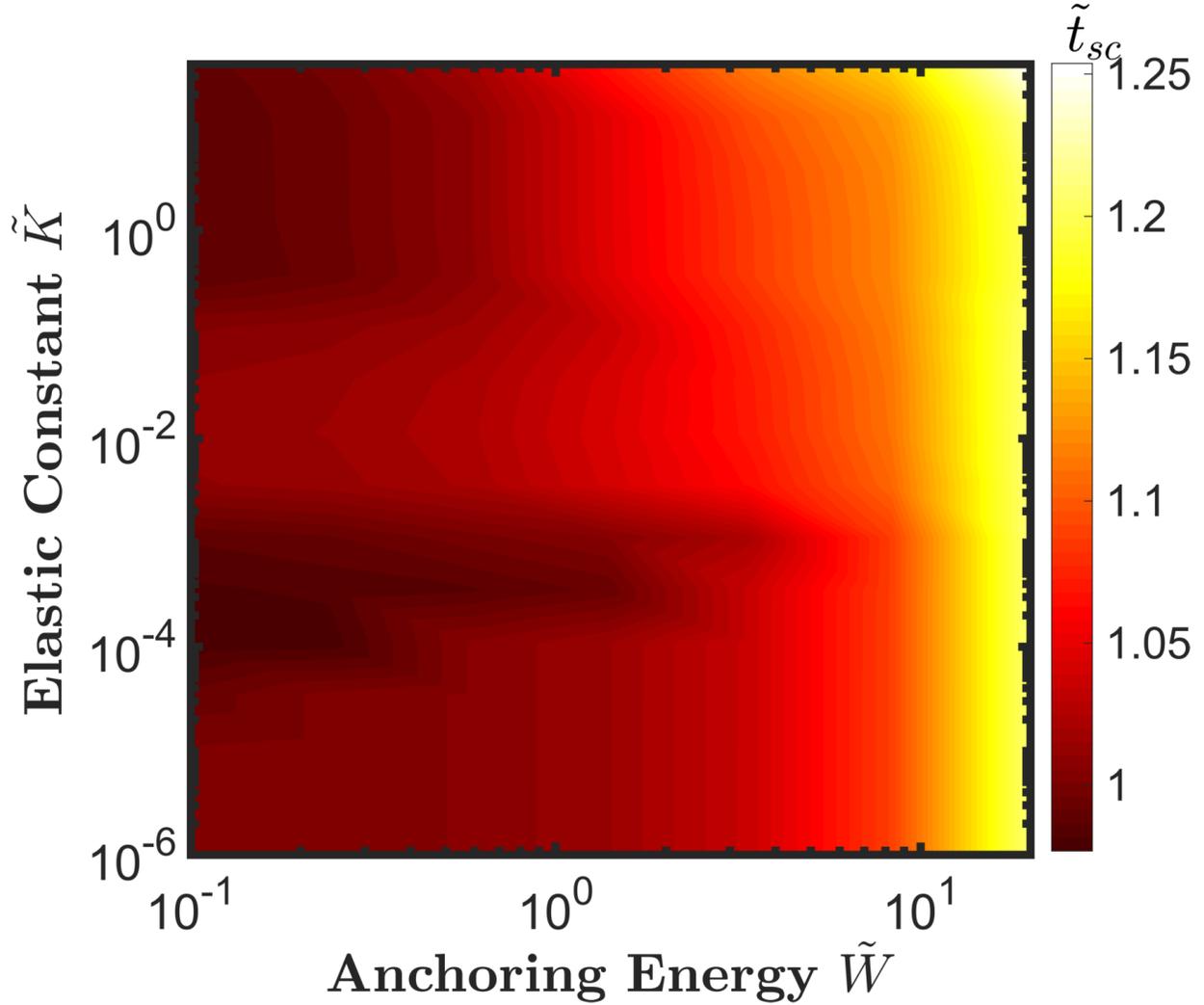

Figure S8: Effect of elastic constant $K$ and anchoring strength $W$ on the dendrite suppression metrics: short circuit time $\tilde{t}_{SC}$. The anchoring strength affects the dendrite suppression capability of the liquid crystalline electrolyte much more than the elastic constant. The short circuit time is normalized using the corresponding time obtained for the standard electrolyte. Note that the ideal short circuit time is not infinite and can only be increased to its maximum value obtained when the electrodeposition is perfectly smooth with RF=0 $\mu$m.

14

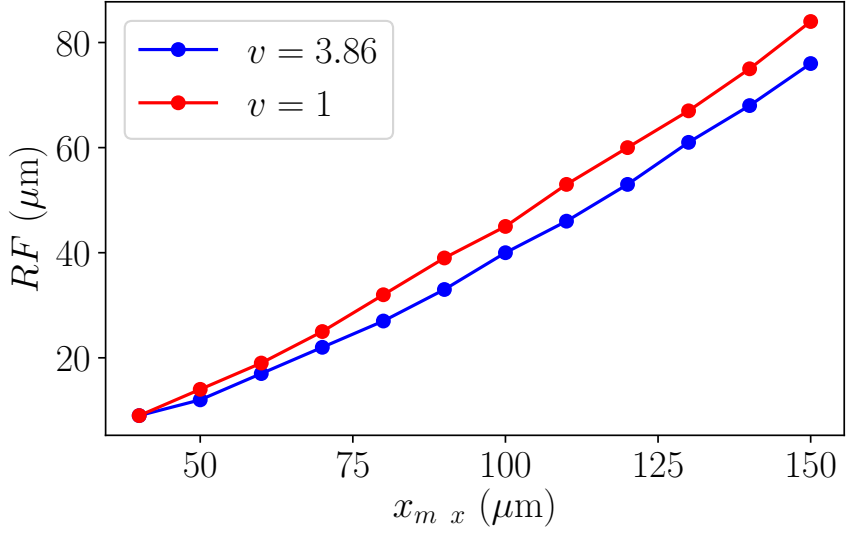

(a)

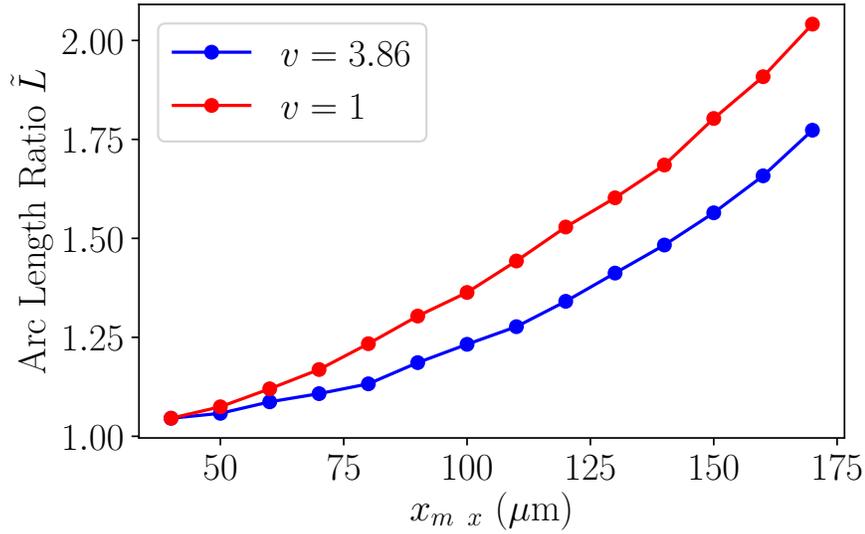

(b)

Figure S9: Variation of (a) roughness factor and (b) arc length ratio $\tilde{L}$ with the maximum x-coordinate of the metal surface plotted for two different values of molar volume of lithium in the electrolyte. $v = V_{Li^+}/V_{Li}$ is the molar volume ratio. Higher molar volume in the electrolyte results in lower roughness and lower arc length ratio. The initial state has a 5 $\mu$m radius hemispherical nucleus.

# Adsorption Energy of Liquid Crystals from First-Principles

First-principles DFT calculations were performed using real-space projector augmented wave method[20] as implemented in GPAW.[21,22] A real space grid spacing of 0.18 Å was used. The substrate used for the calculations were: Si (001) surface (same as the one used in Ref.[19]) and Li (001) surface of similar areas ($\approx$ 230 Å$^2$) and containing four layers. The liquid crystal molecules were placed in $\langle 1\,1\,0 \rangle$ orientation which was found to be the stable orientation along contact with Si (001) surface in Ref.[19] A vacuum spacing of 15 Å was used for the DFT calculations of molecules and 10 Å on either side in the z direction for slab calculations. All ionic coordinates except the bottom two layers of the substrate were completely relaxed until the forces were less then 0.01 eV/Å.

From the DFT calculations of 5CB liquid crystal on Si (001) surface, the adsorption energy per unit area was found to be -0.009 J/m$^2$. The adsorption energies of liquid crystals with Li metal are an order of magnitude higher than this value.



\end{document}